\documentclass{svjour3}                    
\usepackage{graphicx}
\usepackage{mathptmx}      
\usepackage{amsmath}

\DeclareMathSizes{10}{10}{12}{9}

\journalname{Cryogenics}
\begin{document}

\title{Cryogenic broadband vibration measurement on a cryogen-free dilution refrigerator}


\titlerunning{Vibration measurement on a dry dilution refrigerator}        

\author{D.~Schmoranzer \and A.~Luck \and E. Collin \and A.~Fefferman}

\authorrunning{D. Schmoranzer \and A.~Luck \and E. Collin \and A. Fefferman} 

\institute{D. Schmoranzer \and A.~Luck \and E. Collin \and A. Fefferman \at
              Universit\'{e} Grenoble Alpes, CNRS Institut N\'{e}el, 25 rue des Martyrs, BP166, 38042 Grenoble cedex 9, France}

\date{Received: date / Accepted: date}

\maketitle

\begin{abstract}
This manuscript reports a set of acceleration measurements in the frequency range from 0~to~50~kHz performed at the mixing chamber plate (2 axes) and the top flange (3 axes) of a cryogen-free dilution refrigerator. Various configurations of the support frame and coupling to the pulse tube compressor and motor have been tested, and the dominant contribution in the spectrum of vibrations is located, surprisingly, near 20~kHz. Finally, the efficiency of various precautions in suppressing the observed vibration levels is illustrated.

\end{abstract}

\section{Introduction}
\label{intro}

Today, cryogen-free dilution refrigerators (DR) are rapidly becoming the dominant technology for achieving mK temperatures, as they eliminate the need for on-site helium liquefaction and are easier to operate than traditional dilution units. The pre-cooling traditionally provided by a bath of liquid helium and a 1~K pot is replaced by a 2-stage pulse tube cooler, capable of reaching temperatures around 3~K at its cold end. 

Some of the drawbacks of cryogen-free DRs compared to traditional ones relevant to ultra-sensitive measurements at very low temperatures include increased vibration levels~\cite{PT1,PT2} and the lack of a purely cryogenic vacuum space. In this manuscript, the vibration levels at the top flange and mixing chamber plate of such a dilution refrigerator in various configurations are measured and analysed, seeking optimum performance for the purposes of our measurements with sensitive double paddle oscillators~\cite{ourDPO} and development of a continuous nuclear demagnetization refrigerator~\cite{ourCNDR}. Our measurements extend the frequency range investigated in previous work~\cite{BlueforsNoise,DDRNoise} up to 50~kHz.

\begin{figure}[tb!]
\includegraphics[width=0.9\linewidth]{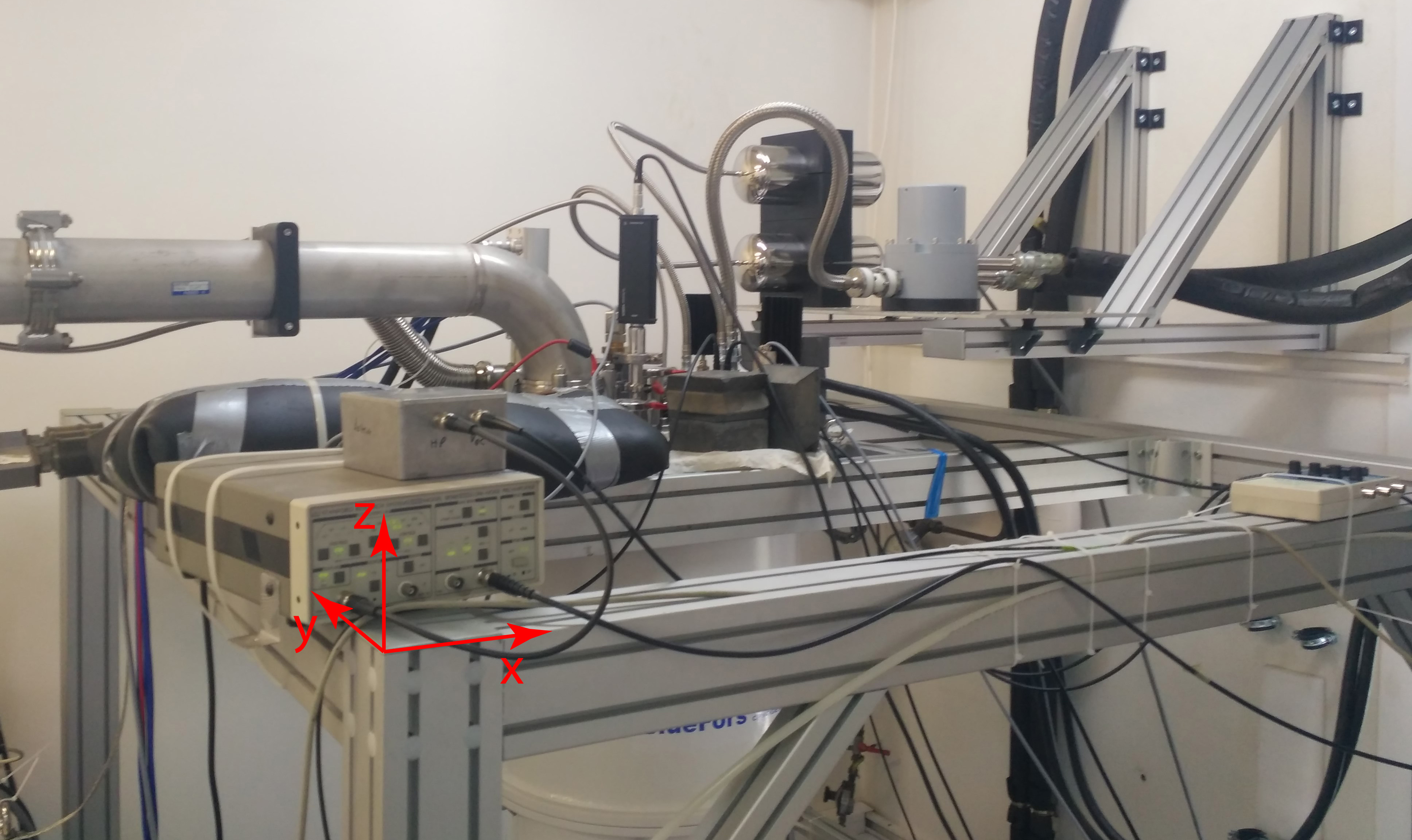}
\includegraphics[width=0.9\linewidth]{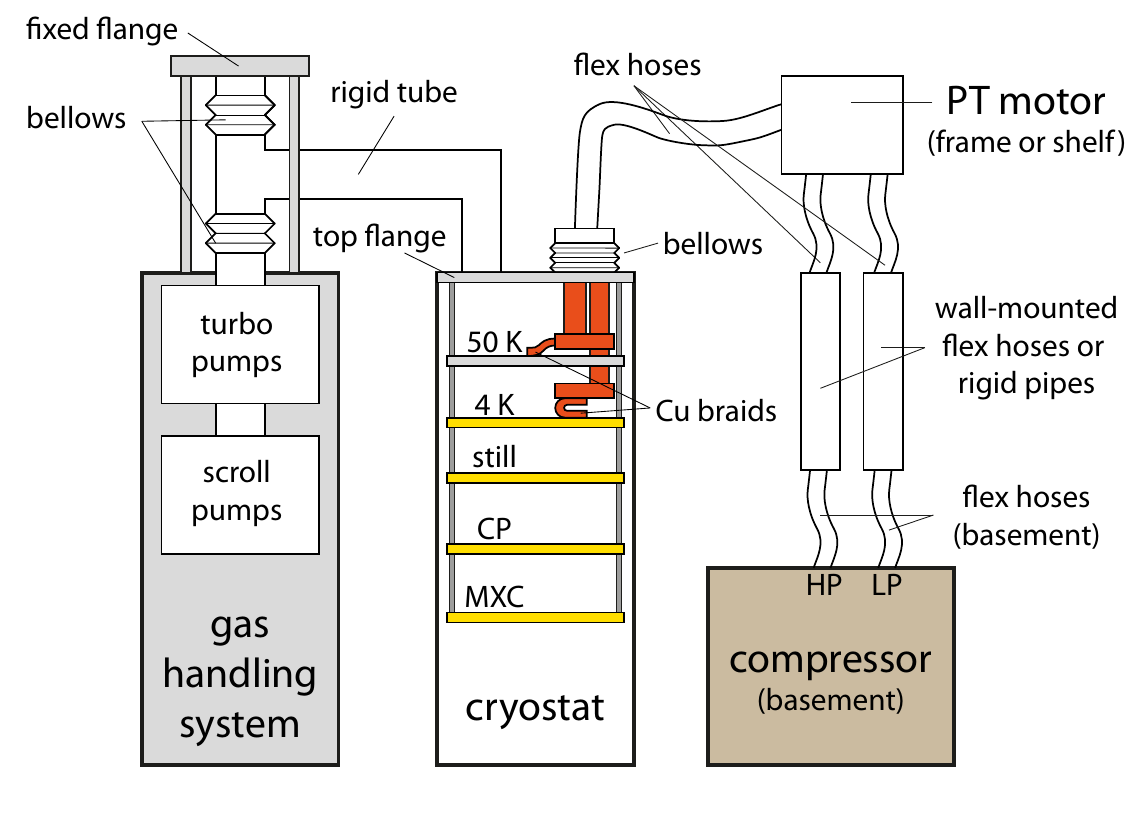}
\caption{Top: a photograph of the top of the cryostat support frame with a sandbag and several lead bricks. The pulse tube motor is placed on a wall-mounted shelf. Arrows mark the reference frame for the acceleration measurements. Bottom: A schematic diagram of the entire setup. The top flange of the cryostat is rigidly connected to the anodized aluminum frame. The compressor unit is located in the basement directly below the laboratory. The gas handling system (not in the photo) contains all pumps necessary to operate the dilution refrigerator.}
\label{fig:setup}       
\end{figure}

\section{Experimental setup}
\label{sec:setup}

Our setup consists of a BlueFors LD-400 dilution refrigerator using the Cryomech PT-415RM cryocooler with a motor model no. RM604152504 and a CPA1110 compressor unit. The refrigerator was purchased with the pulse tube bellows damping stage, but without the optional separate stand for the pulse tube motor offered by BlueFors. The cryostat is mounted on a aluminum support frame fitted with air legs, and connected to the gas handling system placed in the same room via a 100~mm rigid pipe (still pumping line), along with some flexible hoses. The compressor is installed in the basement below the laboratory.

During the measurements, different ways of loading the top flange of the cryostat were tested, including 25~kg sandbags and an assortment of lead bricks with a total weight close to 50~kg. The pulse tube motor, mounted on a 6~mm duraluminum plate, which was originally attached to the cryostat frame via rubber silencer blocks (provided by manufacturer), was later moved to a wall-mounted shelf to test whether this improves the observed vibration levels. Moreover, the original flexible hoses connecting the motor to the compressor were replaced with solid pipes attached to the walls (except for short sections near the pulse tube motor or the compressor unit), and these were later isolated with foam to reduce possible acoustic loading. Both versions of the piping were equipped with AeroQuip fittings, allowing easy manipulation at the full pressure of 250~psi. The setup is shown in Fig.~\ref{fig:setup} along with its schematic diagram.

A three-axis dynamic piezo-accelerometer built at the Institut N\'{e}el from three commercially available accelerometers was used. Two of these were PCB 351B41 by PCB Piezotronics \footnote{http://www.pcb.com/Products.aspx?m=351B41 , accessed 14/09/2018} (used on x- and z-axis) with a nominal frequency range of 1-2000~Hz, and the third was Kistler Model no. 8703A50M8 \footnote{https://www.kistler.com/?type=669\&fid=73055\&model=document , accessed 14/09/2018} (y-axis), with a frequency range of 1-5000~Hz. A sensitive battery-powered amplifier PCB 480B21 with a gain setting of 10 was used to condition the signals, which were then detected by a National Instruments PCI-6052E DAQ card with 100~kHz sampling rate. The accelerometer calibrations provided by the manufacturers are 10.2~mV/(m s$^{-2}$) for PCB 351B41 and 100~mV/g = 10.19~mV/(m s$^{-2}$) for the Kistler accelerometer. The sensor calibrations are valid at room temperature only, and thus in most of this work, only statistical and spectral properties of the acquired voltage signals are presented. The calibrations are, however, applicable to the measurements performed at the top flange as presented below, if the signal bandwidth is limited to the calibrated range.

It should be emphasized that spectra in the frequency range up to 50~kHz are shown and interpreted, exceeding the nominal specifications of the sensors. The purpose is to try and estimate qualitative changes between individual configurations of the refrigerator, with no need for precise calibration or exact numerical values. The sensors are physically capable of detecting vibrations up to 50~kHz, and distinguish between various situations, as shown in the next Section. The intrinsic noise of our detection setup can be characterised as white noise, with a measurable 1/f contribution at frequencies below $\approx 1$~Hz (c.f. Fig.~\ref{fig:MXCspectra} below). The noise level seems to be determined primarily by the amplifier, as for both PCB and Kistler accelerometers, white noise voltage of $u_{w} \simeq 3.5 \times 10^{-5}$~V~Hz$^{-1/2}$ is observed, corresponding to an acceleration noise of $a_{w} \simeq 3.5 \times 10^{-4}$~g~Hz$^{-1/2}$ in the calibrated frequency range at room temperature. It should also be noted that the observed noise level varies to some extent with the charge level of the amplifier battery. The noise levels of our setup are higher than those reported in Ref.~\cite{DDRNoise}, where the same types of sensors have been tested as well and rejected in favour of more sensitive probes.

\section{Results and discussion}
\label{sec:exp}

Two series of measurements were performed: the first one monitoring the vibrations of the top flange at room temperature and the second one with the sensors placed at the mixing chamber plate, which was kept at temperatures close to 4~K.

\subsection{Top flange}
\label{ssec:TF}

\begin{figure}[tb!]
\includegraphics[width=0.99\linewidth]{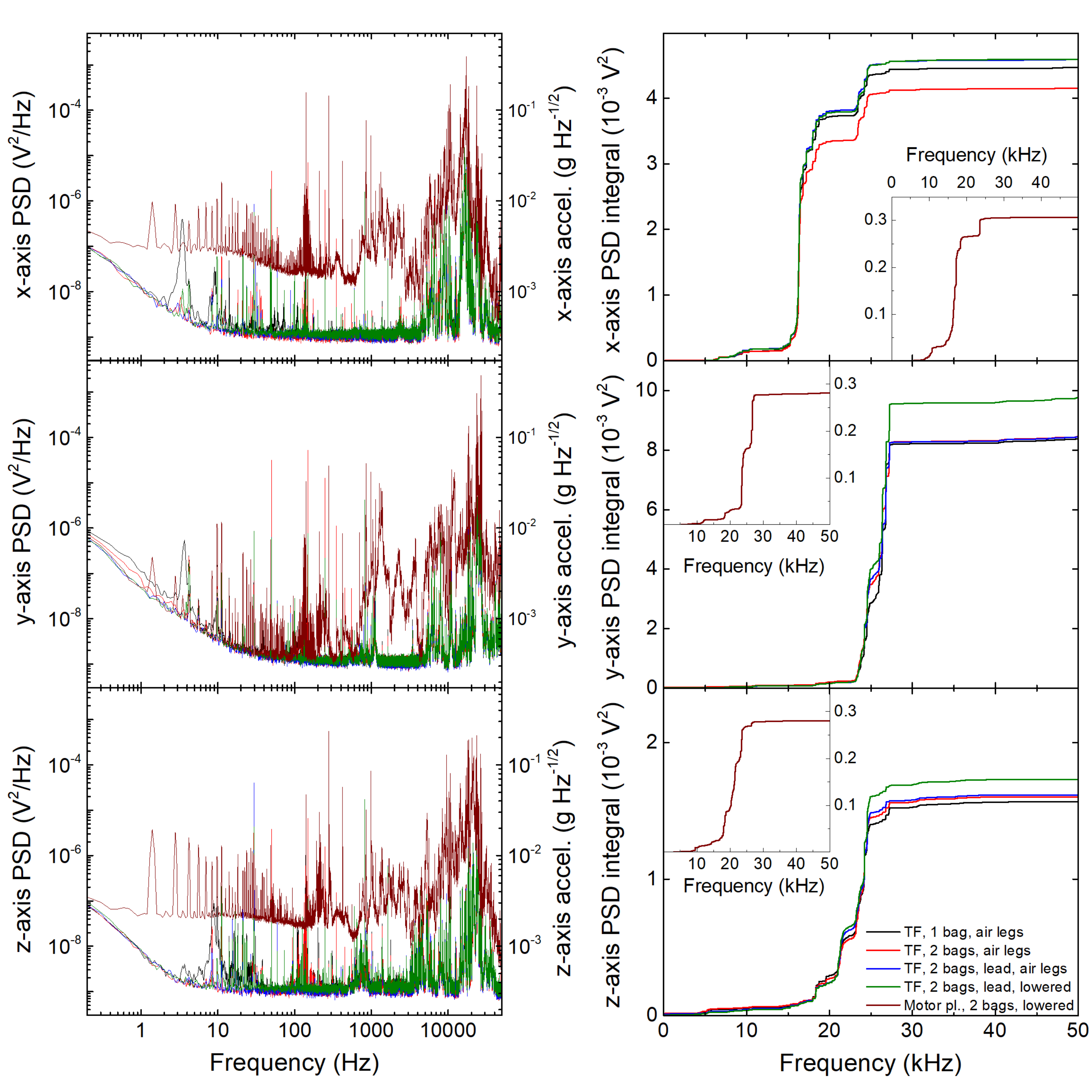}
\caption{Left: Power spectral densities (PSDs) for voltage signals proportional to acceleration in x, y, and z-axes measured on the \textit{top flange (TF)} or the \textit{motor plate}, with loading of the top flange as indicated and the support frame either supported by \textit{air legs} or \textit{lowered} (see legend). Each spectrum shown is obtained by averaging the spectra of 100 measurements of 10~s duration with 100~kHz sampling rate. The right hand y-axes show a conversion to acceleration spectra, which is \emph{applicable only in the calibrated frequency range of each sensor}. Right: Integrals of the same power spectral densities on the top flange (main panels) and on the motor plate (insets, same units), with the background withdrawn. For loading the support frame, up to two \textit{sandbags} of 25~kg were used, together with a set of \textit{lead bricks} of varied size and mass, totaling about 50~kg. The PSDs measured at the top flange differ very little among each other, but it is clearly shown that the vibrations originate largely from the motor plate. Note that when the pulse tube is off, the signal corresponds to the background level of $\approx 10^{-9}$ V$^2$/Hz except for a few isolated peaks, see Fig.~\ref{fig:MXCspectra}, where such a measurement is shown. RMS values of acceleration obtained for each case from a single recorded signal are shown in Tab.~\ref{tab:RMS}.}
\label{fig:topflange}       
\end{figure}

The spectra obtained with the accelerometers at the top flange with different loading of the support frame are shown in Fig.~\ref{fig:topflange} and RMS values of accelerations obtained directly from the recorded signals using the calibrations provided by the manufacturers are listed in Table~\ref{tab:RMS}. In all configurations, one obtains very similar vibration spectra, with only minor differences revealed by integrating the power spectral densities (PSDs). It is, however, clearly shown that most of the vibrations originate from the plate with the pulse tube motor, and that the silencer blocks are insufficient in isolating these vibrations from the cryostat and its frame. Although the kinetic energy is concentrated in the low frequency part of the spectrum, as expected, a substantial contribution to the accelerations is found in closely spaced resonant peaks between approximately 7~kHz and 28~kHz. The vibration spectra acquired at the top flange correspond to accelerations an order of magnitude higher than reported in Ref.~\cite{DDRNoise} on the mixing chamber plate of another brand of cryostat. It is possible that this is due to enhanced acoustic coupling to the noisy environment on the outside of the top flange with respect to sensors located in vacuum on the mixing chamber plate, or due to a more powerful cryocooler model.

\begin{table}[ht!]
	\centering
		\renewcommand{\arraystretch}{1.3}
		\begin{tabular}{c|c c c c c c}
		Configuration & $a_x^{2kHz}$ & $a_y^{5kHz}$ & $a_z^{2kHz}$ & $a'_x$ & $a'_y$ & $a'_z$\\
		 & [\textit{ms}$^{-2}$] & [\textit{ms}$^{-2}$] & [\textit{ms}$^{-2}$] & [\textit{ms}$^{-2}$] & [\textit{ms}$^{-2}$] & [\textit{ms}$^{-2}$]\\
		\hline
			top flange, 1 sandbag, air legs & 0.009 & 0.042 & 0.012 & 0.66 & 0.86 & 0.39\\
			top flange, 2 sandbags, air legs & 0.018 & 0.060 & 0.042 & 0.64 & 0.96 & 0.40\\
			top flange, 2 sandbags+lead, air legs & 0.008 & 0.032 & 0.036 & 0.68 & 0.92 & 0.40\\
			top flange, 2 sandbags+lead, lowered & 0.010 & 0.032 & 0.027 & 0.67 & 0.96 & 0.41\\
			motor plate, 2 sandbags+lead, lowered & 0.23 & 0.24 & 0.21 & 5.65 & 5.19 & 5.32\\
		\end{tabular}
\caption{RMS values of accelerations acquired in each case at the top flange or motor plate directly from a single recorded signal of 10~s duration and 100~kHz sampling rate. The accelerations in the left three columns were obtained using only the calibrated range of the sensors (the indicated bandwidth). The dashed accelerations in the rightmost three columns were obtained from the full frequency range 0.1~Hz - 50~kHz and may be affected by non-linear response of the sensors outside the calibrated frequency range. These values are shown merely for their relative comparison and \emph{should not be regarded as numerically accurate}. In both cases, the calibrations provided by the manufacturers were used to convert voltages to acceleration values, see Section~\ref{sec:setup} for details.}
\label{tab:RMS}
\end{table}

\subsection{Mixing chamber plate}
\label{ssec:MXC}

\begin{figure}[tb]
\includegraphics[width=0.99\linewidth]{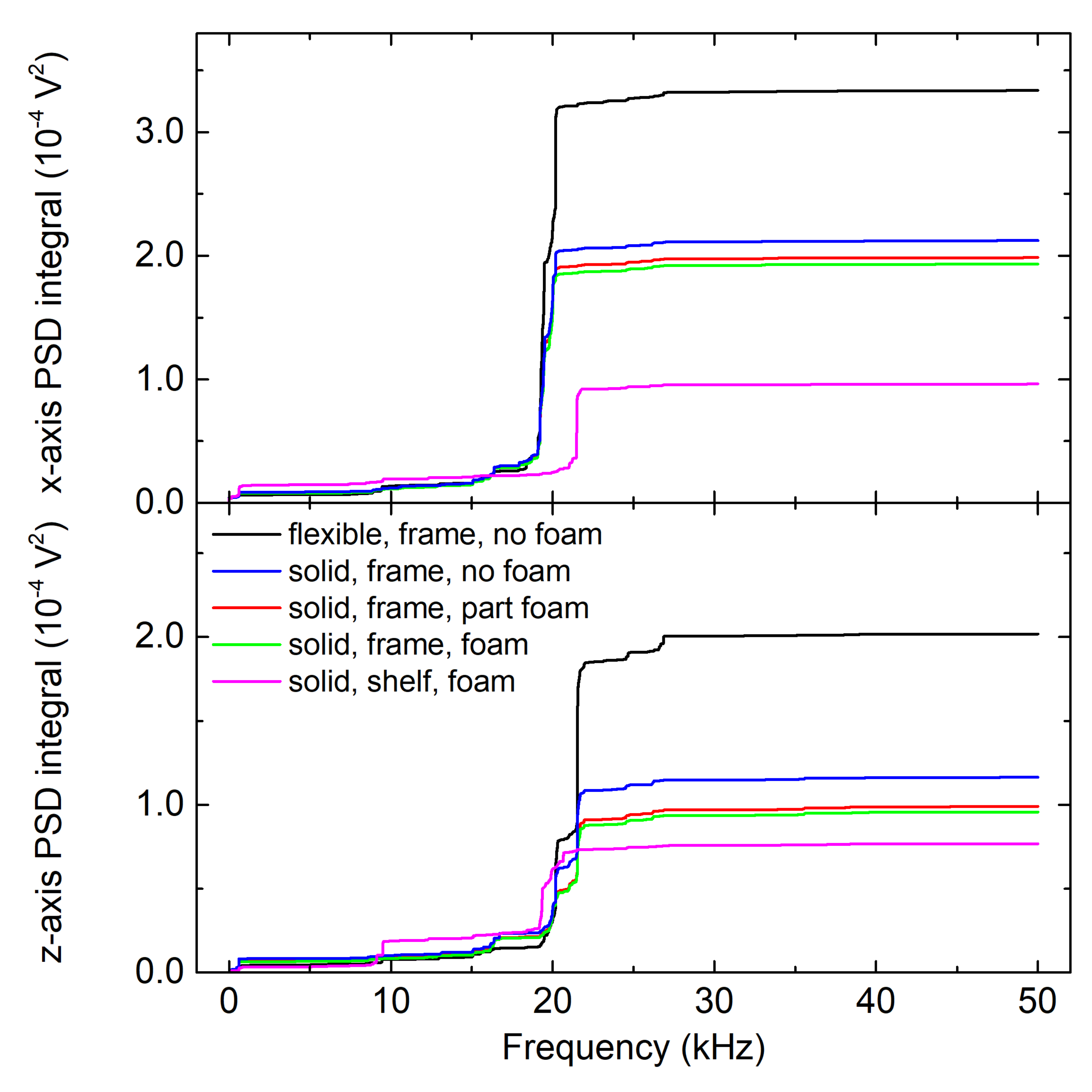}
\caption{Integrated PSDs corresponding to acceleration in the x- (top), and z-axes (bottom), measured at the mixing chamber plate near 4~K at the conditions indicated. The pulse tube motor was positioned on a metal plate mounted via silencer blocks either to the cryostat \textit{frame}, or a wall-mounted \textit{shelf} decoupled from the cryostat frame (see Fig.~\ref{fig:setup}). The motor was connected to the compressor either entirely by \textit{flexible} hoses with stainless steel braiding, or \textit{solid} stainless steel pipes with a short flexible section near the motor. The solid pipes were subsequently covered with isolating \textit{foam} to test the possibility of acoustic transmission of vibrations through air, see text. The background level was determined for each spectrum and subtracted, prior to integration. It is clearly shown that the main contribution to the observed accelerations is at frequencies between 7~kHz and 28 kHz.}
\label{fig:MXC}       
\end{figure}

Next, the sensors were placed at the mixing chamber (MXC) plate and connected to the top flange with specialized cryogenic coaxial cables designed to suppress charge induction when subject to motion. The cryostat cabling permitted the use of two sensors only, hence only the two PCB 351B41 sensors on x- and z-axes were kept. Since the main contribution to the vibrations was found to originate from the pulse tube motor, our focus was on testing the mounting of the motor plate and on the hoses/pipes connecting the motor to the compressor. The results of our measurements are presented in Fig.~\ref{fig:MXC} in the form of integrated PSDs, clearly demonstrating that decoupling the motor from the cryostat frame significantly improves the vibration levels, as does replacing the original flexible hoses connecting the motor to the compressor with solid wall-mounted pipes. The low temperature measurements were made at mixing chamber temperatures ranging from 4.0 to 5.3 K. Measurements at 4.6~and~5.3~K under otherwise identical conditions demonstrated a negligible temperature dependence of the accelerometer response.

\begin{figure}[tb]
\includegraphics[width=0.99\linewidth]{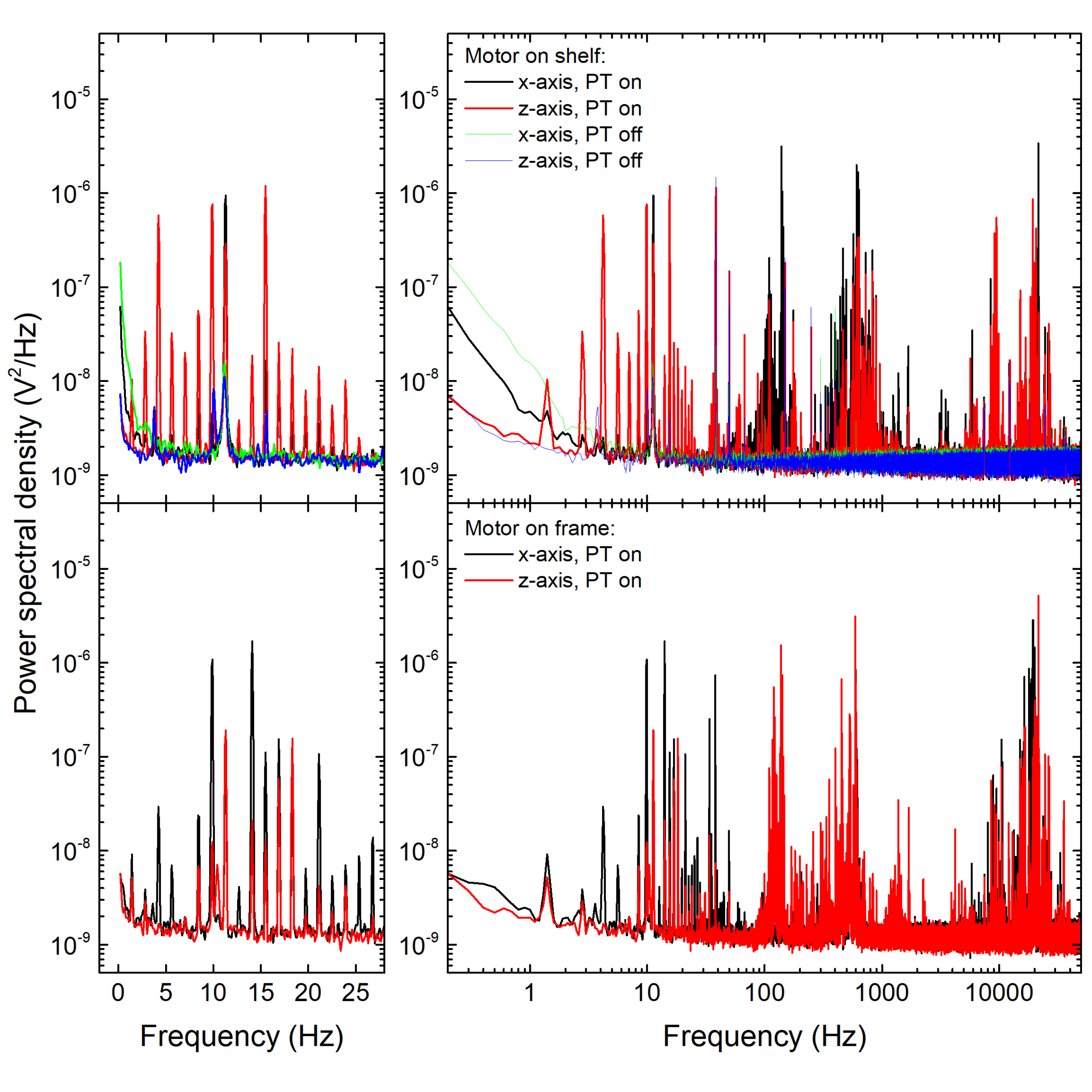}
\caption{PSDs corresponding to acceleration in the x- and z-axes as measured at the mixing chamber plate with solid pipes connecting the pulse tube motor to the compressor, and the motor attached either to the cryostat frame (bottom) or a wall-mounted shelf (top). Peaks corresponding to the multiples of the pulse tube compressor frequency (1.4~Hz) are clearly visible in the left panels. The reader should again be aware that the presented frequency range exceeds the sensor specifications. However, when one compares the spectra obtained with the pulse tube on and off (top panel), it is evident that the sensors are indeed capable of detecting the vibrations around 20~kHz originating from the pulse tube motor.}
\label{fig:MXCspectra}       
\end{figure}

Selected spectra in the form of PSDs are presented in Fig.~\ref{fig:MXCspectra}, highlighting the low-frequency region below 30~Hz, where the pulse tube compressor frequency of 1.4~Hz and its multiples can be observed. However, the dominant contribution is found at frequencies between 7~kHz and 28~kHz, as can be seen from the integrated spectra in Fig.~\ref{fig:MXC}. Moving the motor from the cryostat frame to the shelf had little net effect at low frequencies, except in changing to some extent the direction of the vibrations, aligning them more with the x-axis. At high frequencies around 20~kHz, moving the motor away from the frame reduced the overall vibration levels, as seen again from the integrated PSDs in Fig.~\ref{fig:MXC}. Note that the spectra may contain intrinsic resonances of the sensors, and resonances were indeed observed with the pulse tube off at 7.4, 12.3, 24.5 and 36.9~kHz. Although their origin cannot be identified reliably, it was verified that their contribution to the integrated PSDs is negligible compared to the dominant acoustic contribution between 7~kHz and 28~kHz present when the pulse tube is turned on.

As the high-frequency vibrations are still present on the MXC plate, it is clear that they can propagate from the top flange along the solid supports inside the vacuum can of the cryostat, a known effect discussed in the literature, see e.g.~ Ref.~\cite{DDRNoise}. Since the entire vibration spectrum is, in fact, dominated by vibrations at these frequencies, a question arises, how efficiently can acoustic coupling through air transmit the vibrations from the compressor or from the connecting pipes to the cryostat. 

An additional test was thus carried out, where the pipes were first partly and then completely covered in isolating foam\footnote{Noma Gum rubber foam, internal diameter 27~mm, thickness 13~mm}, showing each time a slight improvement in the integrated PSDs (c.f. Fig.~\ref{fig:MXC}) as well as reducing to some extent the audible noise in the laboratory. This result indeed points at the significance of acoustic transmission of vibrations through air, perhaps warranting sound-proofing of the cryostat for measurements of devices operating at frequencies between 7~and~28~kHz.

\section{Conclusions}
\label{sec:conclude}

Vibration measurements on a cryogen-free dilution refrigerator were performed and analysed, showing a dominant contribution between 7~kHz and 28~kHz, and pointing to several practical measures that lead to the improvement of vibration levels. Using wall-mounted solid pipes to connect the pulse tube motor to the compressor resulted in a significant suppression of vibrations, as did moving the pulse tube motor away from the cryostat frame. Covering the pipes with soft isolating foam led to an additional improvement, albeit less significant than either of the above. On the other hand, loading the support frame with sandbags or lead bricks had very little effect, and led primarily to redistribution of the vibrations in the frequency domain rather than to their suppression.

Further developments related to suppression of vibration level on our cryostat would likely include attaching the large pumping line connecting the still of the DR to the turbomolecular pump(s) in the gas handling systems to the building walls, or by softening its mechanical connection to the cryostat by means of an additional short section of flexible bellows tubing placed horizontally. We hope that our work will stimulate further investigations into the issues of acoustic and ultrasonic vibrations on dry dilution refrigerators.

\begin{acknowledgements}
We would like to thank P.-E. Roche for helpful discussions and for providing us with the accelerometers used in this work. We also thank O. Tissot for his valuable technical assistance. We acknowledge support from the ERC StG grant UNIGLASS No. 714692 and from ERC CoG grant ULT-NEMS No. 647917.
\end{acknowledgements}



\end{document}